\documentclass[prb,preprint,amsmath,superscriptaddress]{revtex4}
\usepackage{amssymb}

\usepackage{amsmath}
\usepackage{graphicx}
\usepackage{bm}

\begin{document}

\title{Coherent spin dynamics in quantum wells in quantizing magnetic
field}
\author{E. Ya. Sherman and J.E. Sipe}
\affiliation{Department of Physics and Institute for Optical Sciences, 
University of Toronto, 60 St. George Street, Toronto, Ontario, Canada M5S 1A7}

\begin{abstract}
We investigate theoretically the coherent longitudinal and transversal spin
relaxation of photoexcited electrons in quantum wells in quantized magnetic
fields. We find the relaxation time for typical quantum well parameters 
between 10$^2$ and 10$^3$ ps. For a realistic random potential the
relaxation process depends on the electron energy and $g$-factor, demonstrating
oscillations in the spin polarization accompanying the spin relaxation. The dependence
of spin relaxation on applied field, and thus on the corresponding "magnetic" length,
can be used to characterize the spatial scale of disorder in quantum wells. 
\end{abstract}

\maketitle

\section{\protect\smallskip Introduction}

A magnetic field applied to a two-dimensional (2D) electron system changes
both the orbital and spin dynamics of the carriers. A crucial aspect of
this dynamics is spin relaxation, which arises due to spin-orbit (SO)
coupling. This coupling is well-described in zincblende
(001)-grown structures by a Hamiltonian $H_{\mathrm{SO}}$ that is the sum of two terms, 
a Rashba Hamiltonian \cite{Rashba84,Rashba60}
${H}_{\mathrm{R}}=\alpha_{\mathrm{R}}\left(\sigma_{x}p_{y}-\sigma_{y}p_{x}\right)/\hbar$ 
and a Dresselhaus Hamiltonian\cite{Dyakonov86,Zutic04} 
${H}_{\mathrm{D}}=\alpha_{\mathrm{D}}\left(\sigma _{x}p_{x}-\sigma_{y}p_{y}\right)/\hbar$, 
where $\alpha _{\mathrm{R}}$
and $\alpha _{\mathrm{D}}$ are coupling constants, $\sigma $ are the Pauli
matrices, and $\mathbf{p}_{\Vert }=\left( p_{x},p_{y}\right) =-i\hbar \nabla
_{\parallel }-(e/c)\mathbf{A}_{\parallel }$.  Here $e$ is the electron charge, and $\mathbf{A}_{\parallel }$ is
a vector-potential of the external field; ${H}_{\mathrm{R}}$ and ${H}_{%
\mathrm{D}}$ arise, respectively, due to the artificial macroscopic asymmetry of the
structure, and due to the microscopic inversion asymmetry of the zincblende
unit cell. The coupling constants $\alpha_{\mathrm{R}}$ and $%
\alpha _{\mathrm{D}}$ typically range from 10$^{-10}$ to 10$^{-9}$ eVcm. 
\cite{alphas}

The spin relaxation of conducting electrons is usually described using
a Dyakonov-Perel' \cite{Dyakonov72} approach, where it is assumed that the
orientation of the spin precession axis changes randomly through scattering
by impurities. In the absence of an external magnetic field the spin
relaxation rate is $\gamma _{\mathrm{DP}}\approx \left( \alpha /\hbar
\right) ^{2}p^{2}\tau $, with $\tau $ the momentum relaxation time,
and where $\alpha $ depends on $\alpha _{\mathrm{R}}$ and $\alpha _{\mathrm{D%
}}.$  Since spin relaxation arises due to the random spatial motion of the
electron, the orbital effect of a magnetic field, which can restrict the
region over which an electron can sample the effect of impurities,
influences the spin relaxation. At $\omega _{c}\tau \gg 1,$ the relaxation
rate decreases by a factor\cite{Ivchenko73} of $\omega _{c}^{2}\tau ^{2}$,
where $\omega _{c}=|e|B/mc$ is the cyclotron frequency for a magnetic field $B$, 
with $m$ being the electron effective mass. In such a strong field the
electron path becomes close to a circle, with scattering 
effects negligible. On this orbit the mean spin
precession angle vanishes due to the fact that the electron velocity 
$\mathbf{v}_{\Vert}(t+\pi /\omega_{c})=-\mathbf{v}_{\Vert}(t)$, and, 
therefore, the randomness in the precession is
suppressed. For a random SO
coupling a non-quantizing magnetic field can, however, speed up spin
relaxation and make the relaxation process Gaussian rather than exponential in its time
dependence.\cite{Glazov05}
The $B-$dependence of the relaxation rate in nonquantizing fields can demonstrate
magnetoquantum oscillations, as shown by self-consistent Born approximation 
for 2D electron gas in a short-range potential of impurities.\cite{Burkov04}  Analysis
of spin relaxation in weak magnetic fields allows the extraction of SO
coupling parameters from experimental data.\cite{Glazov04} 

Studies of spin dynamics of itinerant electrons typically assume that their motion 
can be described semiclassically.
However, spin dynamics of carriers with quantized lateral motion is of interest both for understanding 
of the fundamental physics of spin transport \cite{Mireles02} and for applications of nanosize systems in spintronics. 
An interesting example of a system with quantum lateral motion is 2D electron gas in a strong magnetic
field, where Landau states must be used to represent the electrons and only a few (or a fraction) of the
Landau levels is occupied.
An analogous regime has been widely discussed for quantum dots, where
electrons are confined by an external potential. There the SO coupling mixes
states with opposite spins, making spin-flip transitions
accompanied by phonon emission possible.\cite{dots} 
Here we are interested in itinerant two-dimensional electrons,
where early experimental data has shown the suppression of spin
relaxation in strong magnetic fields.\cite{Potemski89} The treatment of 2D
electrons in quantizing fields is rather subtle,\cite{Ando} with different
theoretical techniques giving different results even for static properties
such as the density of states. The analytical approaches require
approximations that might be not widely applicable. The situation becomes
even more complicated for the response functions, such as those describing
charge and spin currents, as well as relaxation processes. An analytical
study by Bastard \cite{Bastard92} of spin relaxation in the self-consistent Born approximation, 
for a short-range random potential and a small electron $g$-factor, where
the Zeeman splitting is small in comparison with the level broadening,
showed that both spin-orbit coupling and disorder play a role, and allowed an 
estimation of the relaxation rate. 

Here we perform
a numerical study of the problem using the exact diagonalization technique
for a large but finite-size system, and show the diverse physical
mechanisms that contribute to the relaxation process. This approach 
has proven its applicability in calculations of the spin-Hall conductivity of a 
disordered 2D electron gas,\cite{Nomura05} where the spectrum and the full 
set of eigenstates are required for the calculation.

\section{\protect\smallskip Model of disorder and spin dynamics}

We consider a quantum well in a magnetic field parallel to the growth
direction, $\mathbf{B}=B\hat{z}$. The field forms spin-split Landau levels,
inhomogeneously broadened by the disorder. In undoped quantum
wells the random potential arises due to the monolayer islands of the parent
compounds at the interfaces (thickness fluctuations) and due to the content
variations near the interfaces.\cite{Leosson00,Patane95,Castella98} The
island patterns depend on the growth conditions. The Hamiltonian for an
electron in an undoped quantum well with disorder is 
\begin{equation}
H=\frac{{p}_{\Vert }^{2}}{2m} + 
\mu_{B}g\left({\bm \sigma}\cdot\mathbf{B}\right)/2 +
H_{\mathrm{SO}}\left({\bm \sigma},\mathbf{p}_{\Vert }\right) +
U\left(\bm{\rho}\right) ,
\end{equation}
where the random contribution $U\left( \bm{\rho}\right) =\sum_{d}V\left( %
\bm{\rho}-\mathbf{R}_{d}\right) ,$ where $\mathbf{R}_{d}$ is the position of
a defect. As a model, we consider the Gaussian potential $V\left( \bm{\rho }%
\right) =V_{r}\exp \left( -\rho ^{2}/R_{g}^{2}\right) $ with the
areal density of defects $N\left( \bm{\rho}\right) =\sum_{d}\delta \left( %
\bm{\rho }-\mathbf{R}_{d}\right),$ and a correlation function 
$\left\langle N\left( \mathbf{0}\right) N\left( \bm{\rho}\right)
\right\rangle =N_{\mathrm{imp}}\delta \left( \bm{\rho}\right),$ 
where $N_{\mathrm{imp}}$ is the average areal density of defects. 
To avoid
the uniform shift of the Landau levels, we assume that $V_{r}$ $=\pm V_{g}$
varies from site to site, being either positive or negative such that the
mean value $\left\langle U\left(\bm{\rho}\right)\right\rangle=0.$ The
correlation function of the random potential is
$\left\langle U\left(\mathbf{0}\right)U\left(\bm{\rho}\right)
\right\rangle\equiv\left\langle U^{2}\right\rangle F_{c}(\rho)$, where \cite{Efros89} 
\begin{equation}
\left\langle U^{2}\right\rangle =
\frac{\pi }{2}N_{\mathrm{imp}}R_{g}^{2}V_{g}^{2},
\qquad F_{c}(\rho )=e^{-\rho ^{2}/2R_{g}^{2}}.  
\end{equation}

At $B=0$ the momentum relaxation time in this model is given by:
\begin{equation}
\frac{1}{\tau}=N_{\mathrm{imp}}V_{g}^{2}R_{g}\frac{m}{\hbar^{3}}\left\{ 
\begin{array}{ccl}
2\pi ^{2}R_{g}^{3}, &  & \lambda \gg R_{g}, \\ 
&  &  \\ 
\lambda ^{3}/16\sqrt{2}\pi^{3}, &  & R_{g}\gg \lambda ,
\end{array}
\right. \text{ }
\end{equation}
where $\lambda =2\pi/k$ is the electron wavelength. The result for the 
$\lambda\gg R_{g}$ case is valid in the Born approximation for scattering by
well-separated impurities;\cite{Ando} in the opposite limit the electron is moving
semiclassically in a smooth potential, where $\tau\propto\lambda^{-3}$ 
depends on the electron energy.\cite{Wilke}

To describe the magnetic field, we choose the Landau gauge 
$\mathbf{A}=\left(0,Bx,0\right)$ where the eigenstates $\left|nKs\right\rangle$ of the unperturbed
Hamiltonian $p_{\Vert }^{2}/2m+\mu_{B}g\left({\bm\sigma}\cdot\mathbf{B}\right)/2$ are
represented by the spinors:
\begin{equation}
\phi_{nKs}({\bm \rho}) =\frac{e^{ik_{y}y}}{\sqrt{L_{y}}}\frac{1}{\pi
^{1/4}\sqrt{2^{n}n!l_{B}}}\exp \left[ -\frac{x_{K}^{2}}{2l_{B}^{2}}\right]
H_{n}\left( \frac{x_{K}}{l_{B}}\right) \beta _{s},
\end{equation}
where $\beta_{s}$ is the spinor corresponding to one of the states $%
\left| s\right\rangle =\left| \uparrow \right\rangle $,$\left| \downarrow
\right\rangle $, $n$ is the Landau level number, the magnetic length $l_{B}=%
\sqrt{\hbar c/|e|B}$, $x_{K}\equiv x-X_{K}$, $X_{K}=-k_{y}l_{B}^{2}$ is the
center of the oscillator wavefunction, $k_{y}=-2\pi K/L_{y},$ $K=0,1,\ldots
,K_{\max },$ $L_{y}$ is the $y-$axis size of the system and $H_{n}$ is the $n$th
Hermite polynomial. The corresponding unperturbed spectrum is $E(n,s)=\hbar \omega
_{c}(n+1/2)\pm g\mu_{B}B/2.$ Due to the selection rules for the matrix
elements of $\mathbf{p}_{\Vert}$ and spin components $\sigma_{x}$ and $\sigma_{y}$, 
the SO coupling only connects states of opposite spins
from nearest Landau levels.\cite{Schliemann03} This results in a small shift
in the energies of the order of $m\left(\alpha/\hbar\right)^{2}$. The matrix
elements of the disorder Hamiltonian diagonal over the spin index are given
by:
\begin{equation}
H_{\mathrm{rnd}}(n^{\prime }K^{\prime }s^{\prime};nKs)=
\int d^{2}\rho\overline{\phi}_{n^{\prime}K^{\prime}s^{\prime }}({\bm \rho})
U\left( \bm{\rho}\right)
\phi_{nKs}({\bm \rho}),
\end{equation}
and couple states in all Landau levels for which $\left| X_{K}-X_{K^{\prime }}\right| \lesssim
l_{B},$ and, correspondingly $\left|k_{y}-k_{y}^{\prime }\right| \lesssim
1/l_{B}$, thus leading to electron localization both in the $x-$ and $y-$%
directions. As a result the density of eigenstates with
energies $E_{j}$ has the form of broadened Landau levels,
typically (at $B=5$ T) with the width $\Gamma$ on the order of or less than 1 meV, 
at least an order of magnitude smaller than $\hbar\omega_{c}$. 
The corresponding eigenfunctions are expressed as
linear combinations:

\begin{equation}
\left|\psi_{j}\right\rangle
=\sum_{n,K,s}a_{nKs}^{(j)}\left| nKs\right\rangle,
\end{equation}
with complex coefficients $a_{nKs}^{(j)}$. Only a few of the
amplitudes $a_{nKs}^{(j)}$ are nonnegligible, since $U\left(\bm{\rho}\right)$ 
couples only the states with close momenta $k_{y}.$
Due to the combined effect of disorder and SO coupling, 
the $\left|nK_{1}\uparrow \right\rangle $ and $\left|nK_{2}\downarrow\right\rangle$ 
states within one Landau level become coupled as
shown in Fig. 1(a), thus introducing randomness in the spin precession at the
frequency scale $\Gamma/\hbar$, and, in turn, spin relaxation. 

We consider an electron-hole plasma
injected by a light pulse in an undoped quantum well (Fig.1(b)). 
We assume that the plasma density is small, and neglect all many-body effects. 
Alternatively, for a doped quantum well, we assume the carriers are injected
into unoccupied Landau states above the Fermi level. 
We concentrate on the spin relaxation of electrons, since the hole spins relax much faster.\cite{Hilton02}
 
We assume that the spectral width of the exciting light $\Delta\omega$
satisfies the conditions $\Gamma\ll\hbar\Delta\omega\ll\hbar\omega_{c}$ as shown in Fig.1(b).
Therefore,
the states $\left|\psi^{\rm in}_i(t=0)\right\rangle$ (index $1\le i\le N_{\rm in}$ numerates the injected 
electrons, with their total number being $N_{\rm in}$) in which the electrons are injected 
can be written as superposition of the  $\left|nK_{i}s\right\rangle$ states from essentially
one Landau level $n_{\rm in}$ and spin projection $s_{\rm in}$ determined by the light 
polarization.\cite{Pfalz05} Within this approximation a unitary transformation connects the full
sets of  $\left|\psi^{\rm in}_i(t=0)\right\rangle$  and  
$\left|n_{\rm in}K_{i}s_{\rm in}\right\rangle\equiv \left|\Phi_{K_{i}}(t=0)\right\rangle$ states.
Thus, in place of averaging the time-dependent 
spin components over  $\left|\psi^{\rm in}_i(t)\right\rangle$, 
we can average over the  $\left|\Phi_{K_{i}}(t)\right\rangle$ as follows:
\begin{eqnarray}
&&\left\langle \sigma _{\zeta}(t)\right\rangle =\frac{1}{N_{\rm in}}
\sum_{i=1}^{N_{\rm in}}
\left\langle \psi^{\rm in}_{i}(t)\right|\sigma _{\zeta }\left| \psi^{\rm in}_{i}(t)\right\rangle \\ \nonumber
&=&
\frac{1}{N_{\rm in}}
\sum_{i=1}^{N_{\rm in}}
\left\langle \psi^{\rm in}_{i}(t=0)\right|
e^{iHt/\hbar}\sigma _{\zeta }e^{-iHt/\hbar}\left| 
\psi^{\rm in}_{i}(t=0)\right\rangle \\ \nonumber
&=&
\frac{1}{N_{\rm in}}
\sum_{i=1}^{N_{\rm in}}
\left\langle \Phi^{\rm in}_{K_{i}}(t)\right|\sigma _{\zeta }\left|\Phi^{\rm in}_{K_{i}}(t)\right\rangle. 
\end{eqnarray}
This relaxation is coherent in the sense that 
the total energy $\left\langle H\right\rangle$
of the injected electrons is conserved on the timescale considered here  
and is not transformed into 
lattice phonons or low-energy electron excitations.

To evaluate $\left\langle\sigma_{\zeta}(t)\right\rangle$ numerically,
we have to find the spinor representation of the $\left|\Phi^{\rm in}_{K_{i}}(t)\right\rangle$;
that is, we must determine the eigenstates of the Hamiltonian $H$ in Eq.(1). To do this,  
we have chosen a basis of $n_{B}=256$ states per Landau level for each spin projection. The
total Hilbert space included $N_{L}=6$ Landau levels, which we find sufficient 
for our choice of the parameters given below, for which $\hbar\omega_{c}\gg\Gamma$. 
The coefficients $a^{j}_{nKs}$ considered as the elements of the
eigenvectors are arranged in the following order: 
$\left(\left\{a_{0K\uparrow}\right\},\left\{a_{0K\downarrow}\right\},
\ldots,\left\{a_{N_{L}-1K\uparrow}\right\},\left\{a_{N_{L}-1K\downarrow }\right\}
\right),$ where in every subset $K$ is running from 0 to $K_{\max}.$

\section{\protect\smallskip Numerical results: the role of the $g$-factor and the disorder landscape}

For sample calculation we consider two types of
typical structures. The first is a symmetric quantum well with electrons 
located in a GaAs layer, with the
Dresselhaus SO coupling $\alpha_{\mathrm{D}}=0.35\times 10^{-9}$ eVcm, 
$m=0.067m_0$ ($m_0$ is the free electron mass), and $g=-0.45$. The
other is an asymmetric structure with the electrons located in a In$_{0.5}$Ga$_{0.5}$As layer, 
with\cite{Nitta97} a Rashba SO coupling $\alpha_{\mathrm{R}}=0.35\times 10^{-9}$
eVcm, $m=0.05m_0$, and $g=4$. We assume that both of them have a width of 10 nm. For these structures the ratio of effective masses 
$m(\mathrm{GaAs)}/m(\mathrm{In}_{0.5}\mathrm{Ga}_{0.5}\mathrm{As})=1.35$, while the electron $g-$factors are different approximately by a factor 
of $|g(\mathrm{In}_{0.5}\mathrm{Ga}_{0.5}\mathrm{As})/g(\mathrm{GaAs})|=8.9$. 
It is the difference in $g$-factors that will lead to large quantitative 
difference between the spin relaxation in these two structures.

As a realization of the disorder\cite{Patane95} we consider a short-range potential with $V_{g}=3.5$
meV for GaAs and  $V_{g}=5.0$ meV for $\mathrm{In}_{0.5}\mathrm{Ga}_{0.5}\mathrm{As}$
structure, $R_{g}=3$ nm, and $N_{\mathrm{imp}}=10^{12}$ cm$^{-2}$. These
parameters can describe the random potential arising due to the single unit cell layer
thickness variations in quantum wells of the width of 10 nm. Since $l_{B}\gg R_{g}$,
the width of the Landau level $\Gamma=\hbar\sqrt{2\omega_{c}/\pi\tau}$ does not depend on the level number.
We also consider a long-range potential with $R_{g}=20$ nm and the same amplitudes $V_{g}$
as for the short-range one. 
Both these random potentials lead to mobilities of $\approx 5\times 10^{4}$ cm$%
^{2}/$Vs at concentrations of electrons $N_{\mathrm{el}}\approx 5\times 10^{11}$
cm$^{-2}$ in both structures. Note that in the case of a long-range
potential $R_{g}\gg l_{B}$, we have $\Gamma =2\sqrt{\left\langle U^{2}\right\rangle}$.
This spatial scale can be achieved by quantum well
fabrication with growth interruption \cite{Leosson00}, or by remote doping on the quantum well sides.

The results of numerical calculations for the GaAs quantum well are presented in
Figs. 2 and 3 for different initial spin components $\left\langle\sigma_{z}(0)\right\rangle =-1$ 
and $\left\langle\sigma_{x}(0)\right\rangle =1$,
respectively, two different types of disorder, and magnetic fields of 4 T and 8 T. 
In these GaAs-based structures we find the relaxation times on the order of 100 ps,
of the same order of magnitude for out-of-plane $\left\langle\sigma _{z}(t)\right\rangle$ and in-plane 
$\left\langle\sigma_{x}(t)\right\rangle$ spin components.  They relax on the same 
time scale since the same mechanism, that is the random spin precession, leads to the relaxation in both
spin polarizations. With the increase of the Landau level number, the electron motion becomes
less sensitive to the disorder, and the spin relaxation time increases, as demonstrated by 
the results for $n=2$ in Fig. 2 and Fig.3. This corresponds well
to the experimental results of Sih {\it et al.}\cite{Sih04} Spin relaxation
due to acoustic phonon emission \cite{Frenkel91} occurs at a much longer
time scale, and is not considered here. An interesting effect is the spin precession in the 
$\left\langle\sigma_{z}(t)\right\rangle$ relaxation, clearly seen in our
calculations. The spin-orbit interaction couples Zeeman-split Landau levels
which are broadened due to disorder, and so the 
$z$-component of the spin is not a constant of motion.  
For this reason, the initially prepared  $\sigma_{z}=\pm 1$ states precess with the frequencies 
in the range determined by the Zeeman splitting and the Landau level width, distributed 
over interval of the width $\Gamma/\hbar$ centered 
at $g\mu_{\rm B}B/\hbar$. The precession amplitude is smeared with time due to spin relaxation, as 
seen in our results.

Fig. 4 presents the results for In$_{0.5}$Ga$_{0.5}$As structure,
where the spin relaxation is much slower. Due to the increased Zeeman splitting
the frequency of the spin oscillations here is considerably larger than that  
in the GaAs quantum well, and the oscillations become more well-defined.
The role of the Zeeman splitting $|g|\mu_{B}B$ in
spin relaxation, clearly seen when one compares the results for the GaAs and
In$_{0.5}$Ga$_{0.5}$As structures, is crucial to the mechanism of the spin
relaxation, and can be understood as follows.\cite{gfactor} The random potential
accompanied by SO coupling leads to the spin-flip transitions. However, only
spatially close states (with a large overlap) with opposite spins can
contribute effectively to the spin relaxation. On the other hand, the spin-flip 
process should conserve energy,
and, therefore, a lateral
distance $\ell_{\rm s}$ on which the orbital has to be displaced to find its spin-flip 
partner state depends on the Zeeman
splitting. To understand the effect we evaluate the fluctuation of the
expectation value of energy for a state described by wave function $\psi\left(\bm{\rho}\right)$: 
\begin{widetext}
\begin{eqnarray}
\left\langle
\left(\Delta U_{\bm{\rho}}\right)^{2}
\right\rangle &=&
\left\langle 
\left[ 
\int U
\left(\bm{\rho}_{1}\right) 
\left(
\psi^{2}\left(\bm{\rho}_{1}\right)-
\psi^{2}\left(\bm{\rho}_{1}+\bm{\rho}\right)
\right)
d^{2}\rho_{1}
\right]^{2}
\right\rangle \\ \nonumber
&=& 2\langle U^{2}\rangle
\int F_{c}(\bm{\rho}_{1}-\bm{\rho}_{2})
\psi^{2}\left(\bm{\rho}_{1}\right)
\left[
\psi^{2}\left(\bm{\rho }_{2}\right)-
\psi^{2}\left(\bm{\rho}_{2}+\bm{\rho}\right) 
\right] 
d^{2}\rho_{1}d^{2}\rho_{2}.
\end{eqnarray}
\end{widetext}

This fluctuation can be calculated in two limiting cases as:
\begin{equation}
\left\langle\left(\Delta U_{\bm{\rho}}\right)^{2}\right\rangle
=2\left\langle U^{2}\right\rangle
\left\{ 
\begin{array}{l}
1-F_{c}(\rho ),\quad l_{B}\ll R_{g}, \\ 
\\ 
{\displaystyle{\frac{R_{g}^{2}}{l_{B}^{2}}}}
\left[
1-e^{-\rho^{2}/4l_{B}^{2}}
\right]
,\quad l_{B}\gg
R_{g},
\end{array}
\right. 
\end{equation}
where we assumed for an estimate that for the ground orbital state $\psi
^{2}\left( \bm{\rho }\right) =\exp \left( -\rho ^{2}/2l_{B}^{2}\right) /2\pi
l_{B}^{2}.$ If $\ell_{\rm s}$ determined by energy 
conservation is less than $l_{B},$ that is $|g|\mu _{B}B\leq
\left\langle U^{2}\right\rangle^{1/2}\min \left(
l_{B}/R_{g},R_{g}/l_{B}\right),$ the relaxation occurs effectively. In the
opposite case, the decay time increases due to a small spatial overlap of
the initial and final states. In other words, if $|g|\mu _{B}B>\Gamma,$ the
spin relaxation is suppressed, since the energy conservation cannot be
fulfilled in the spin-flip process. In this case the $z$-component of spin relaxes from the 
initial value by approximately  $\Gamma/|g|\mu_{B}B$, and then the system has to
pass by emitting acoustic phonons through a phonon bottleneck \cite{Frenkel91} for the spin to  relax it further.

This "spin-flip distance" argument explains the difference between the spin relaxation for a "long" and "short"- range
potentials shown in Fig. 4. In both cases the amplitude of the potential fluctuations is the same, but in 
the case of a long-range 
potential the electron must be displaced a longer distance to find its spin-flip partner state. Thus, the relaxation 
rate decreases with the increase of the correlation length $R_g$, in agreement with results presented in Fig.4.

\section{\protect\smallskip Conclusions and possible applications}

To conclude, we investigated the spin relaxation in quantizing magnetic
fields in a disordered 2D electron gas, and found that the result depends on
the Landau level number, and that the process is accompanied by spin oscillations. The
mechanism discussed here is different from the usual Dyakonov-Perel' mechanism, and is 
closer in  nature to the Elliot-Yafet mechanism, in which the spin relaxation rate
increases with the disorder.\cite{Zutic04} 
The relaxation rate depends on
the details of the potential and electron $g-$factor, and cannot be understood
solely in terms of the electron mobility.

The results obtained can be used to
characterize the disorder in undoped quantum wells. Such disorder is typically
probed experimentally by studying the inhomogeneous broadening of the spectra of excitons, 
a technique restricted either to the spatial scale given by the exciton Bohr radius 
or by the exciton localization length.\cite{Castella98} 
Neither of these lengths can be changed externally in a well-controlled way. 
The advantage of studying the spin relaxation in a magnetic field
is that it allows probing different spatial scales of disorder by
controllably changing the length $l_{B}$ trough varying the magnetic field. 

\textit{Acknowledgment.} We are grateful to H.M. van Driel, M.E. Flatt\'{e}, A. Najmaie, K. Nomura,
J. Prineas, J. Sinova, and A. Smirl for valuable discussions, and to P. Chak and F. Nastos for
help in numerical calculations. 
This work was supported by the DARPA SpinS program and by the FWF  grant P15220. 

\newpage


\newpage

\begin{figure}[tbp]
\includegraphics[width=6.0cm]{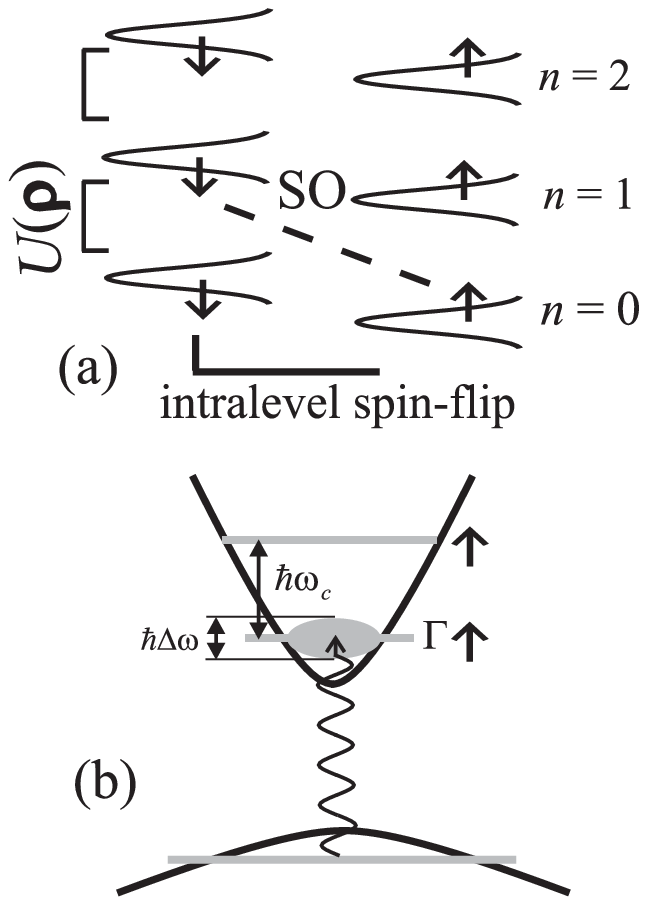}
\caption{(a) Schematic plot of the broadened Landau levels and interlevel transitions for 
$\alpha_{\rm R}=0,\alpha_{\rm D}\ne 0$. Arrows label electron spins, and the index $n$ corresponds to
the Landau level number. (b) Schematic plot of the optical transitions between the Landau levels. Only 
one spin projection is presented.}
\end{figure}
\vspace{2cm}

\newpage

\begin{figure}[tbp]

\includegraphics[width=5.0cm]{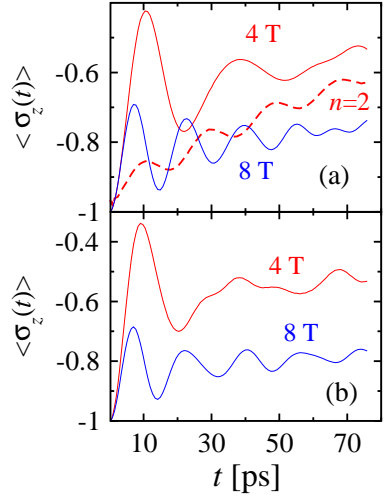}
\caption{(Color online) (a) $\left\langle \sigma _{z}(t)\right\rangle $ for a short-range
($R_{g}=3$ nm) potential, and (b) $\left\langle \sigma _{z}(t)\right\rangle $ for a long-range
($R_{g}=20$ nm) potential in the GaAs quantum well. Solid lines correspond to $n=0$; the dashed
line in Fig.2(a) corresponds to $n=2$, $B=4$T. The strength of magnetic field is shown near 
the lines.}
\vspace{16cm}
\end{figure}

\newpage

\begin{figure}[tbp]
\vspace{6cm}
\includegraphics[width=5.0cm]{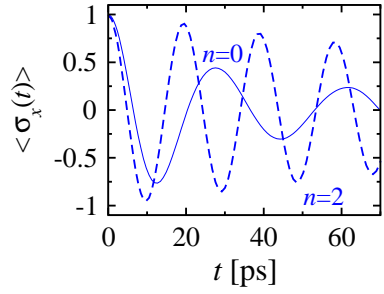}
\vspace{1cm}
\caption{(Color online) $\left\langle \sigma _{x}(t)\right\rangle $ for a short-range
potential ($R_g=3$ nm) in the GaAs quantum well, with $B=4$ T. Landau level numbers 
are shown near the lines.}
\vspace{5cm}
\end{figure}

\newpage

\begin{figure}[tbp]
\includegraphics[width=6.0cm]{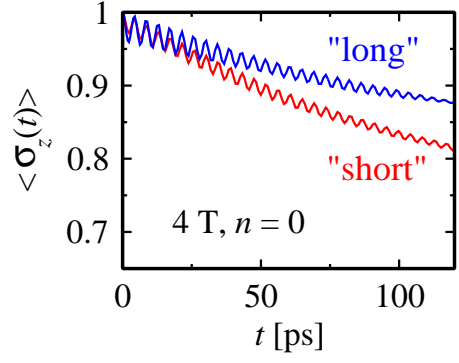}
\caption{(Color online) $\left\langle\sigma_{z}(t)\right\rangle$ for long-range
and short-range potentials in the In$_{0.5}$Ga$_{0.5}$As quantum well.}
\end{figure}


\begin{thebibliography}{99}
\bibitem{gfactor} We mention that a model calculation with $m=0.05m_0$ and $g=0.45$ shows the
results similar to those in Fig.2 but different from Fig.4.
\end{thebibliography}
\end{document}